# O(1) Time Generation of Adjacent Multiset Combinations

## Tadao Takaoka

Department of Computer Science, University of Canterbury
Christchurch, New Zealand
E-mail: tad@cosc.canterbury.ac.nz

**Summary.** We solve the problem of designing an O(1) time algorithm for generating adjacent multiset combinations in a different approach from Walsh [17]. By the word "adjacent", we mean that two adjacent multiset combinations are different at two places by one in their vector forms. Previous O(1) time algorithms for multiset combinations generated non-adjacent multiset combinations. Our algorithm in this paper can be derived from a general framework of combinatorial Gray code, which we characterise to suit our need for combinations and multiset combinations. The central idea is a twisted lexico tree, which is obtained from the lexicographic tree for the given set of combinatorial objects by twisting branches depending on the parity of each node. An iterative algorithm which traverses this tree will generate the given set of combinatorial objects in constant time as well as with a fixed number of changes from the present combinatorial object to the next.

## 1. Introduction

In this paper, we design an O(1) time algorithm for generating adjacent multiset combinations based on a unified approach of twisted lexico tree and tree traversal. Here O(1) means we spend O(1) time from object to object. Roughly speaking, the twisted lexico tree for a set of combinatorial objects can be obtained from the lexicographic tree for the set by twisting branches depending on the parity of each node. We give an even parity to the root. As we traverse the children of a node, we give an even and odd parity alternately to the child nodes. If a node gets even, the branches from the node remain intact. If it gets odd, the branches from the node are arranged in the reverse order. This concept of parity is applied to the entire tree globally, so that the constant change property can be easily shown for each set of combinatorial objects. Although the use of trees and parity in combinatorial generation is found in Zerling[19] and Lucas[10], the tree traversal mechanism in this paper is new. The general concept of this paper can be viewed as a refinement of combinatorial Gray code [4], [14], and [18], which is, in turn, a generalization of the generation of binary reflected Gray code [12]. Previous algorithms [2] and [3] generate multiset combinations in O(1) time in the worst case, but they are not adjacent. Ruskey and Savage [13] left open an even weaker problem of generating those objects in O(1) amortized time. The Gray code for adjacent compositions suggested by Knuth was implemented by Klingsberg [5] in O(1) amortized time. Thus present paper gives a stronger result, as compositions are a special case of multiset combinations. We generate multiset combinations in a one-dimensional array for a vector form. The generation process can be viewed as moving pebbles from box to box one by one, where each box corresponds to a multiset component and has its own size, and there are k pebbles and n boxes. We exhaust all possible arrangements.

This problem is also known as the bounded composition problem in Walsh [17]. That is, given integer k, we exhaust all possibilities (objects) of expressing k by a sum of n non-negative integers $a_i$, $k = \sum a_i$, where each $a_i$ is bounded by a positive integer $b_i$. Walsh generates the composition from the largest lexicographic order in a similar way to Klingsberg, and the value $a_i$ alternates, increasing and decreasing from object to object. This is a complicated algorithm where 23 cases are analyzed to indentify the changing positions. Walsh managed to solve this problem in O(1) time for an object with O(n) extra space. Our program is much shorter at the cost of increased data structures. In a way, we shift the complexity in control structure to that of data structure. The present work is an evolution of the author's previous works on the O(1) time generation of combinations in-place [15] and multi-set permutations [16], in both of which the technique of tree traversal was used. The work in [15] is an improvement on combination generation by Nijenhuis and Wilf [11], Bitner, Ehrlich, and Reingold [1], and Lehmer [8]. The work in [16] is an improvement of Korsh and Lipschutz [6], whose algorithm uses a linked structure rather than one dimensional array for multiset permutations. Korsh and LaFollette [7] achieved O(1) time in-place for the same problem. In-place for combinations means we use O(k) space to generate k-combinations out of n elements in array. It is open whether there is an O(1) time algorithm for generating multiset combinations in place. We partially solve this problem by generating multiset combinations in array of size k in O(1) time using O(n) extra space.

The paper consists of the following sections. In Section 2, we give a recursive framework for generating multiset combinations in lexico-graphic order and Gray code order. In Section 3, we give the concept of twisted lexico tree, which is the main device (date structure) for our generation. In Section 4, we give a generic algorithm for tree traversal as the main control structure. In Section 5, we map the set of multiset combinations onto a twisted lexico tree, and observe a fixed number of changes are enough from object to object. Section 6 gives implementation details. In Section 7, we partially solve the problem of generating multiset combinations in-place. Section 8 concludes the paper. Programs are given in Pascal-like pseudo code. Most types are integer and just a few Boolean.

## 2. Preliminaries with Recursive Framework

Let $m_i$ (i=1, …, n) be the multiplicity of the i-th component in the multiset. that is, there are $m_i$ elements of the i-th component. We express the i-th component by i. We take k elements out of the given multiset and call it a k-combination. Let $(a_1, … , a_n)$ be the vector expression of a multiset combination, where $a_i$ is the repetition of the i-th component. We call the direct expression of components the in-place expression. As mentioned in the last section we first generate multiset combinations in lexico-graphic order.

Note that if $m_i = 1$ for all i, the set of objects becomes that of combinations. If $m_i = k$ for all i, the objects become compositions of k by n integers for $n \leq k$.

**Example 1.** $(m_1, …, m_n) = (1, 2, 2, 1, 1)$, and k=4. We have the following 18 combinations in lexicographic order.

```
Vector form      in-place form
0 0 2 1 1          3 4 4 5
0 1 1 1 1          2 3 4 5
0 1 2 0 1          2 3 3 5
0 1 2 1 0          2 3 3 4
0 2 0 1 1          2 2 4 5
0 2 1 0 1          2 2 3 5
0 2 1 1 0          2 2 3 4
0 2 2 0 0          2 2 3 3
1 0 1 1 1          1 3 4 5
1 0 2 0 1          1 3 3 5
1 0 2 1 0          1 3 3 4
1 1 0 1 1          1 2 4 5
1 1 1 0 1          1 2 3 5
1 1 1 1 0          1 2 3 4
1 1 2 0 0          1 2 3 3
1 2 0 0 1          1 2 2 5
1 2 0 1 0          1 2 2 4
1 2 1 0 0          1 2 2 3
```

The number of combinations can be computed by the inclusion-exclusion principle. Although this is well known, we include an example to confirm the number of multiset combinations. (See Liu [9, p96 ] for example). Let $C(n, k) = n!/(k! (n-k)!)$ be the number of (ordinary) combinations of k elements out of n. Let a multiset A be given by $(m_1, \ldots, m_n)$ such that $m_i$ is between 1 and k. Let B be the k-closure of A, defined by $B = (k, \ldots, k)$. That is, we can take an arbitrary number of elements from each component of the multiset B for k-combinations. Let $S(A, k)$ be the set of k-combinations of A. Let $A_i$ be the set of k-combinations of B which have at least $(m_i + 1)$ elements of the i-th component. Then $|S(A, k)|$ can be computed as

$$|S(A, k)| = | (S(B, k) - A_1) \cap \ldots \cap (S(B, k) - A_n)|$$

$$= |S(B, k)| - \Sigma |A_i| + \Sigma |A_i \cap A_j| + \ldots + (-1)^{-n}|A_1 \cap \ldots \cap A_n|$$

Now observe that $|A_i| = |S(B, k - m_i - 1)|$, since there is a one-to-one correspondence between k-combinations in $A_i$ and k-combinations in $S(B, k - m_i - 1)$; removing $(m_i + 1)$ i's from a k-combination in the former results in one in the latter, and vice versa with adding $(m_i + 1)$ i's to the latter. Similarly we have $|A_i \cap A_j| = |S(B, k - m_i - m_j - 2)|$, etc. To compute $|S(B, k)|$, we have the formula $|S(B, k)| = C(n+k-1, k)$.

**Example 2.** To compute the number of combinations in Example 1, we have
$|S(B, k)| = C(8, 4) = 70$
$|A_1| = |A_4| = |A_5| = C(6, 2) = 15,$ $\qquad |A_2| = |A_3| = C(5, 1) = 5$
$|A_1 \cap A_4| = |A_1 \cap A_5| = |A_4 \cap A_5| = C(5, 0) = 1$
all other terms = 0, and thus
$|S(A, k)| = 70 - (15 + 15 + 15 + 5 + 5) + (1 + 1 + 1) = 18.$

Let us generate multiset combinations in vector form in array a. Let the multiplicity of item i be m[i]. In the following we define b[i]=m[i]+m[i+1]+…+m[n]. Procedure mset generates multiset k-combinations from position i to position n. The lowest possible value of a[i] is maximum of k-b[i+1] and 0. This is because if a[i]+b[i+1]<k, we cannot scatter k balls from box i to box n since b[i+1] is the maximum possible capacity beyond position i and a[i] plus this value cannot make k from position i to n. If this value is negative, 0 must be chosen. The value of upper is similarly reasoned. A recursive procedure mset(i, k) generates multiset k-combinations in array a from position i to n in lexico-graphic order as follows:

```
procedure mset(i, k)
begin
    lower:=max(k-b[i+1], 0); upper:=min(m[i], k);
    if i ≤ n then begin
        for j:=lower to upper do begin
            a[i]:=j;
            mset(i+1, k-j); /* handles remaining k-j balls */
        end
    end
end;
begin {main}
    for i:=n down to 1 do b[i]:=b[i+1]+m[i];
    mset(1, k)
end
```

The recursive algorithm for Gray code order is given next. We have a global array for parity, d, which controls the direction of increasing pattern and decreasing pattern at position i; d[i]=1 for increasing, d[i]=-1 for decreasing. At the end of a call at position i, d is flipped over.

```
procedure mset(i, k)
    var j, lower, upper;
begin
    lower:=max(k-b[i+1], 0); upper:=min(m[i],k);
    if i ≤ n then begin
        if d[i]>0 then
        for j:=lower to upper begin
            a[i]=j;
            mset(i+1, k-j);
        end
        else
        for j:=upper downto lower begin
            a[i:]=j;
            mset(i+1, k-j);
        end;
        d[i]:=-d[i]
    end
    else output;
end;
begin {main}
    for i:=n down to 1 do begin b[i]:=b[i+1]+m[i]; d[i]:=1 end;
    mset(1, k)
end.
```

## 3. Twisted Lexico Tree

The recursive algorithm in the previous section is essentially to traverse a tree along procedure calls. We define such trees in this section. Let $\Sigma = \{\sigma_0, \dots , \sigma_{r-1}\}$ be an alphabet for combinatorial objects. A combinatorial object is a string $a_1 \dots a_n$ of length n such that each $a_i$ is taken from $\Sigma$ and satisfies some property. A total order is defined on $\Sigma$ with $\sigma_i < \sigma_{i+1}$. Let $\Sigma^n$ be the set of all possible strings on $\Sigma$ of length n. The lexicographic order $<$ on $\Sigma^n$ is defined for $\mathbf{a} = a_1 \dots a_n$ and $\mathbf{b} = b_1 \dots b_n$ by

$$\mathbf{a} < \mathbf{b} \Leftrightarrow \exists j \ (1 \le j \le n) \ a_1 = b_1, \dots , a_{j-1} = b_{j-1}, a_j < b_j.$$

Let $S \subseteq \Sigma^n$ be a set of combinatorial objects. The order $<$ on S is defined by projecting the lexicographic order on $\Sigma^n$ onto S. The lexicographic tree, or lexico tree for short, of S is defined in the following way. Each $\mathbf{a} \in S$ corresponds to a path from the root to a leaf. The root is at level 0. If $\mathbf{a} = a_1 \dots a_n$, $a_i$ corresponds to a node at level i. We refer to $a_i$ as label for the node. We sometimes do not distinguish between node and label. If $\mathbf{a}$ and $\mathbf{b}$ share the same prefix of length k, they share the path of length k in the tree. The children of each node are ordered according to the labels of the children. A path from the root to a leaf corresponds to a leaf itself, so $\mathbf{a}$ corresponds to a leaf. The combinatorial objects at the leaves are thus ordered in lexicographic order on S.

The twisted lexico tree of a set S of combinatorial objects is defined as follows together with the parity function. We proceed to twist a given lexico tree from the root to leaves. Let the parity of the root be even. Suppose we processed up to the i-th level. If the parity of a node v at level i is even, we do not twist the branches from v to its children. If the parity of v is odd, we arrange the children of v in reverse order. If we process all nodes at level i, we give parity to the nodes at level i+1 from first to last alternately starting from even. We denote the parity of node v by parity(v). When we process nodes at level i in the following algorithms, which are children of a node v such that parity(v)=p, we say the current parity of level i is p. Note that (labels of) nodes at level i are in increasing order if the parity of the parent if even, or equivalently if the current parity of level i is even. If the parity is odd, they are in decreasing order. We draw trees lying horizontally for notational convenience. We refer to the top child of a node as the first child and the bottom as the last child.

If the labels on the paths from the root to two adjacent leaves in the twisted lexico tree for S are different at nodes not more than a fixed number, we can generate S from object to object with the fixed number of changes, and we say S satisfies the constant change property(CCP). As shown in Section 2, it will be easy to design recursive algorithms that traverse those twisted lexico trees, since they can control the paths to the leaves, and paths back to the calling points.

## 4. Generic Iterative Algorithm

We devise in this section a general framework of iterative algorithm which avoids the O(n) overhead time by recursive calls. Although a similar method is known, this algorithm is new in using data structures "up" and "down", and also with the concept of solution point. The array "up" is to keep track of the position of an ancestor to which the algorithm comes back from an up-point. An up-point is a node which has a single child and from which we go back to an ancestor. The array "down", used in [16], is to keep track of positions to which we go down after we make necessary changes when moving from a node to the next node. Note that this algorithm takes O(1) time in the worst case if S satisfies the CCP and the changing mechanism is properly given. The variable "$v_i$" is for the node to be processed at level i. In many applications, the difference at level i is solved at level j down the tree. We refer to level j as the solution point. These changing positions are called pivots in [17].

**Algorithm 1.** Iterative tree traversal

1. Initialize array a to be the first object in S; Initialize array up and down;
2. Initialize $v_1$, ... , $v_n$ to nodes on the path to the first object (top path);
3. **for** i:=0 **to** n **do** p[i]:=0;    /* all parity values are 0 (even) initially */
4. Initialize i to the first up-point; up[0]:=0;   /* From this point on, i is for the current level */
5. **repeat**
6.     **output**(a);
7.     Perform changes on a at $v_i$ and related positions;
8.     Let $v_i$ go to the next node at level i;  /* Label increasing or decreasing depending on p */
9.     up[i]:=i;
10. **if** $v_i$ is the last child of its parent (largest or smallest label) **then begin**
11.         up[i]:=up[i-1]; up[i-1]:=i-1;   {value of up propagates downwards}
12.         Compute the solution point for up[i];
13.         Update the value of down for up[i];
14.         p[i]:=1-p[i];
15.     **if** $v_i$ is an up-point then i:=up[i] else i:=down[i]
16.     **end**
17. **until** i=0.

The situation is illustrated in the following figure.   Levels of A and B are given by j and k.

          level i=up[j]                level j   k         level n

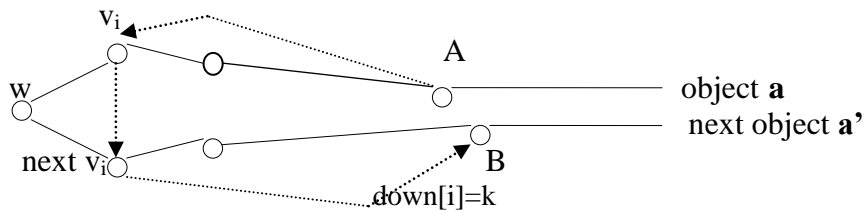

Figure 1. General picture of tree traversal

When we come to the last child of a parent (w in the above figure), we have to update up[i] to up[i-1] so that   we can come back directly from an up-point to w or its ancestor if w itself is a last child. We refer to the paths from $v_i$ to **a** and from next $v_i$ to **a'** as the current path and the opposite path. A current path and opposite path consist of last children and first children respectively except for level i and above. We refer to the points A and B in this figure as an up-point and a landing point. From nodes A and B we have straight lines to the leaves. Node $v_i$ is called the return ancestor of A. Note that the return ancestor is the first non-last child we encounter when we trace the current path from A up towards the root. We also refer to level i as the crossing level when viewed from A and B. How to avoid traversing straight lines is the central problem in this paper. Simply speaking , the algorithm repeats (up, cross, down)-actions.

## 5. Generation of Multiset Combinations

Now we form the twisted lexico-tree for multiset combinations. Here we modify the concept of parity slightly in such a way that we do not give a parity to a node which has a single child. This is slightly different from the parity in Section 3, where every node has parity, because we cannot maintain the parity in straight lines in the tree as we skip them.

**Example 3.** The twisted lexico tree of Example 1 is given below.

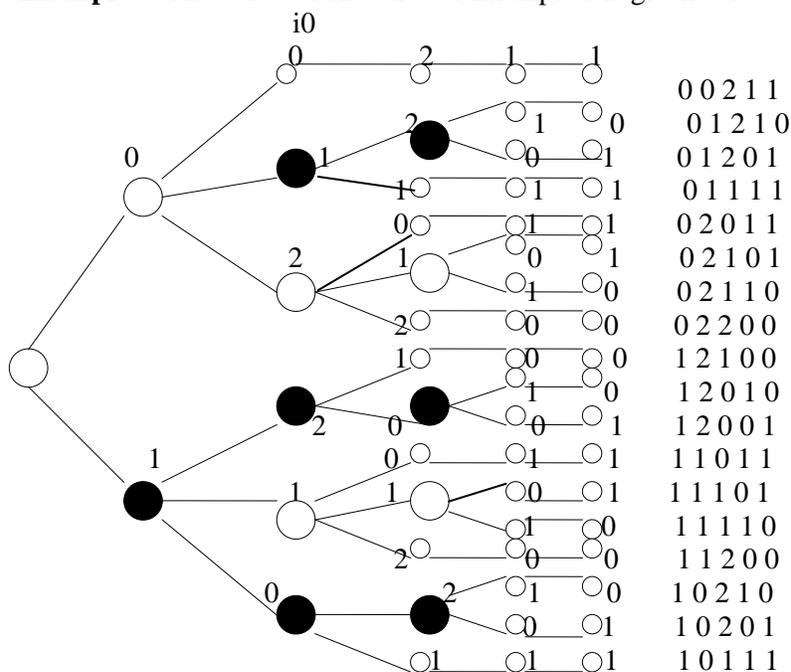

Figure 2. Twisted lexico tree for multiset combinations

In this figure, nodes which are given parities are shown by big circles: white for even and black for odd. We assume even for all nodes on the first path, although some are little circles. Those are regarded as either parity in the proof below.

**Theorem 1.** The set of k-combinations S(A, k) of multiset A given by $(m_1, \ldots, m_n)$ satisfies the constant change property in vector form.

**Proof.**     Form the twisted lexico-tree for the set S(A, k). Let **a** $= a_1 \ldots a_{i-1} a_i \ldots a_n$ and **a'** $= a_1 \ldots a_{i-1} a'_i \ldots a'_n$ be two adjacent k-combinations in the tree. Then $a'_i = a_i + 1$ or $a'_i = a_i - 1$. Along the paths from $a_i$ to $a_n$ and from $a'_i$ to $a'_n$ we observe the difference at i will be solved at some j such that $a'_j = a_j - 1$ or $a_j + 1$ when $a'_i = a_i + 1$ or $a_i - 1$ respectively. All other labels are equal along those paths since we have opposite parities on those paths and the sums of labels on those paths are equal to k.

## 6. Implementation

Now we consider the problem of implementation. The key point in implementation is to prepare necessary information for the opposite path when we are traversing the current path through last children.

We use array indices rather than subscripts for convenience. We identify a node by its label whenever convenient. We maintain the parity at level i by array element d[i]. If d[i]>0, the parity is even and a[i] is increasing, and vice versa. We keep the solution point in array "solve". When we move from a node to the next node at level i, we perform changes (see lines 9 and 10 in the algorithm) by

$$a[i] := a[i] + d[i] \qquad \text{and} \qquad a[j] := a[j] - d[i], \qquad \text{where } j = solve[i].$$

The difficulties are how to find the solution point j and where to go after these changes. If the current node after these changes has a single child, this current node is a last child and we have to go up to the return ancestor, guided by array "up". Otherwise we go down to the node on the opposite path that has a single child, that is, the landing point, guided by array "down".

The computational process is modelled by moving k pebbles, one each time, in n boxes whose i-th box has the capacity of $m_i$. At the beginning we fill the boxes to their capacities from right to left. Let i0 be the right-most non-filled box position. Then we start from i0 and use the first solution point which is set to n initially, changing **a** to **a'** and go down to down[i0] which is set to n-1 initially.

The computation of solution point is based on the array "sum" defined by sum[i] = a[1] + ... + a[i-1], and array "b" defined by   b[i] = m[i] + ... + m[n]. When we stand at a node at level i, we see that the sibling nodes can take the values between "lower" and "upper" where

$$\text{lower} = \max\{k - b[i+1] - sum[i], 0\} \qquad \text{and} \qquad \text{upper} = \min\{k - sum[i], m[i]\}.$$

When d[i]>0 and a[i]=upper, or d[i]<0 and a[i]=lower, we can say we reached a last child at level i. Then we have to prepare the solution point for up[i] which is the level where the return ancestor is. This computation depends on the values of "lower" and "upper" at the next stage on and beyond the opposite path. We name these lower and upper values "lower1" and "upper1" one of which becomes the value of "next" depending on the parity. That is,

$$\text{lower1} = \max\{k - b[i+1] - sum[i] - d[up[i]], 0\} \quad \text{and}$$
$$\text{upper1} = \min\{k - sum[i] - d[up[i]], m[i]\}$$

Let "next" be the next node at level i on the opposite path. If "next" is not equal to a[i], then obviously we can set solve[up[i]] to i. If next=a[i], on the other hand, we set solve[up[i]] to solve[i], the solution point of i. This is because if a[i] is an up-point, we want to recover the sequence $a'_{i+1}$

...a'$_n$ to be a"$_{i+1}$...a"$_n$, where a"$_1$...a"$_n$ is the sequence immediately before **a**. If a[i] is not an up-point, solve[up[i]] will be overwritten later by a[i]'s descendants. See the following figure for illustration and line 18 in Algorithm 2. If a"[i+1] $\neq$ a[i+1], for example, solve[up[i]]=solve[i]=i+1, that is, a'[i+1] becomes equal to a"[i+1] after crossing at level up[i].

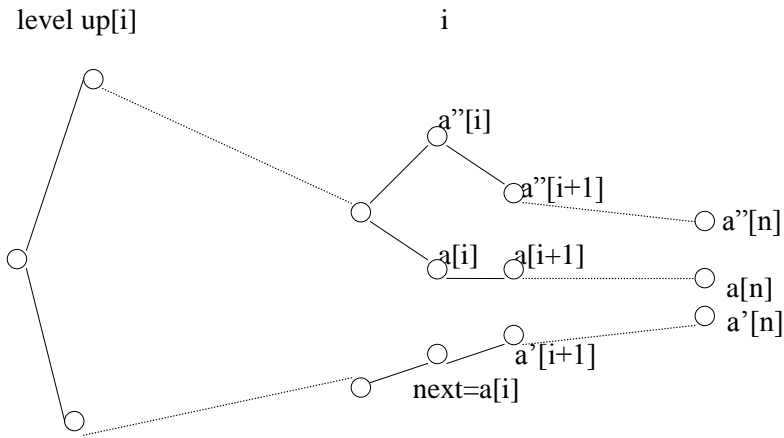

Figure 3.   Illustration for array "solve"

If the landing point on the opposite path is i or closer to the leaf, the information for the solution point is available from the previous path as the above figure illustrates.

If the landing point on the opposite path is closer to the root than the up-point of the current path is, however, the solution point of the landing point has not been set, since we set the solution point for up[i] and the points between up[i] and i are not taken care of. Fortunately, when we come down to the landing point, we can adopt the solution point of the crossing level. This is because we recover the same path to the leaf (path A) after we have the straight line from the landing point (path B). The care for solution points is kept in array "mark". If mark[i] = false, it means the solution point for level i is not prepared. The situation is illustrated in the following figure.

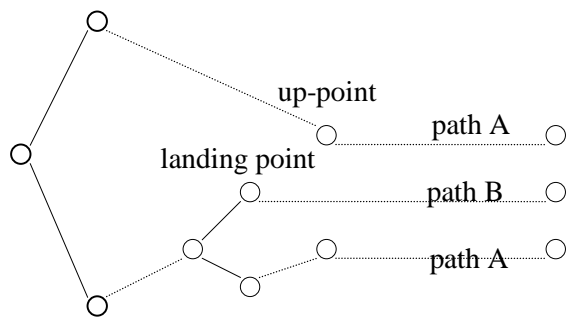

Figure 4. Care of "solve" at landing point

Now we describe how to maintain array "down". If the node "next" at level i has a single child and has one or more siblings, then the node "next" is the landing point on the opposite path, indicated by "next_landing = true". We mark this level by array "up1" by setting up1[i]:=i. Suppose i increased as we went down through the current path. See line 27 in the algorithm. This value of "up1" propagates through last children. With the help of array "up1", we classify the situation into three cases. Dotted lines consist of several branches. Suppose we are standing at "a[i]" in the following figures.

Case 1. If lower1=upper1 ("next" has no sibling), we set down[up1[i]]:=i. This may be overwritten by descendants.

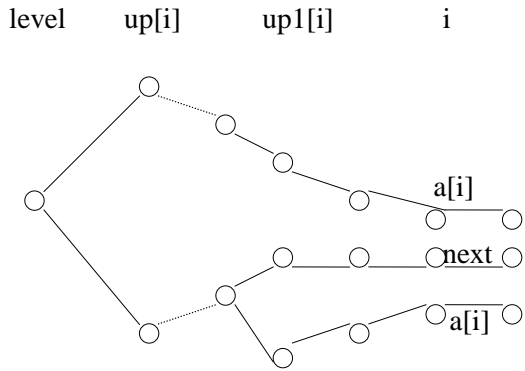

Figure 5. Care of "down" at level up1[i]

When we go up from level i to up[i], the values of "down" for nodes between level i and level up[i] will need care. Fortunately only one value at level up1[i] needs care apart from level up[i].
The correctness of this part and case 3 comes from the fact that the two opposing paths across one or more straight lines are identical in labels from level up1[i]+1 towards leaves, and thus old values of "down" at those levels can be used with no changes.

Case 2. If lower1 ≠ upper1 ("next" has siblings) and "next" is the landing point, we set down[up[i]]:=i.

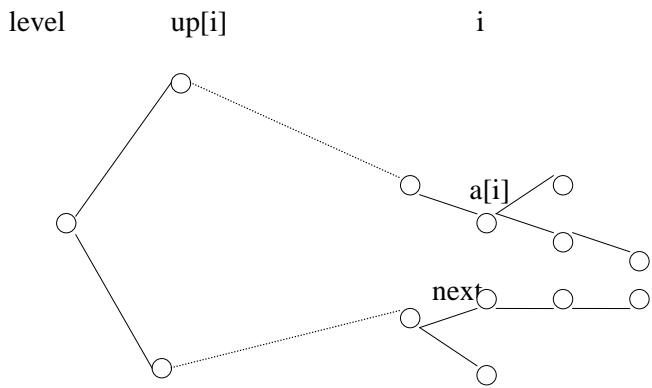

Figure 6. Care of "down" at level up[i] with i

Case 3. If lower1 ≠ upper1 and "next" is not a landing point, we set down[up[i]]:=down[i].
The value of down[i] was prepared using case 1 repeatedly when we traverse from A to B.

level          up[i]                    i          down[i]

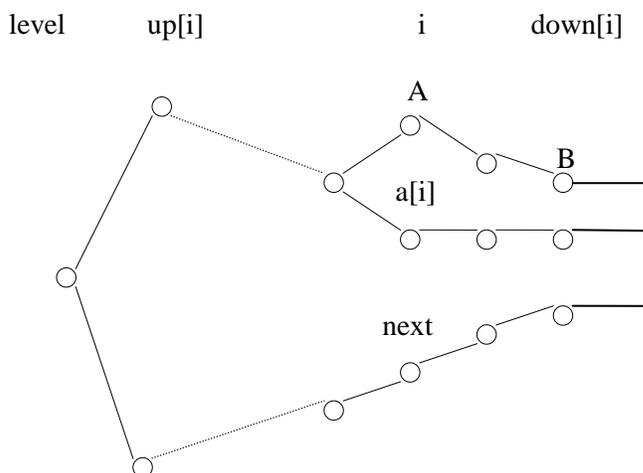

Figure 7. Care of "down" at level up[i] with down[i]

Now the algorithm follows with comments on a few lines.

**Algorithm 2**. Iterative algorithm for generating multiset combinations
1.     Initialize array **a** and sum; Initialize d[i] to 1 for i=1, ..., i0, and -1 for i=i0+1, ..., n;
2.     Initialize down[i] to i-1, up[i], up1[i] to i, solve[i] to n, and mark[i] to **false** for i=1, ..., n;
3.       i :=i0;
4.   **repeat**
5.       **output(a)**;
6.       lower := max{k-b[i+1]-sum[i], 0};
7.       upper := min{k-sum[i], m[i]};
8.       **if not** ((d[i]>0 and a[i]=upper) **or** (d[i]<0 and a[i]=lower)) **then begin**
9.           a[i] := a[i] + d[i];
10.          a[solve[i]] := a[solve[i]] - d[i]
11.      **end**;
12.      up[i] := i;
13.      **if** (a[i]>0 **and** a[i]=upper) **or** (d[i]<0 **and** a[i]=lower) **then begin**
14.          up[i] := up[i-1]; up[i-1] := i-1;   {value of up propagates downwards}
15.          lower1 := max{k-b[i+1]-sum[i]-d[up[i]], 0};
16.          upper1 := min{k-sum[i]-d[up[i]], m[i]};
17.          **if** d[i]>0 **then** next := upper1 **else** next := lower1;
18.          **if** a[i] <> next **then** solve[up[i]] := i **else** solve[up[i]] := solve[i];
19.          mark[up[i]]:= **true**; mark[i]:= **true**;
20.          up_point := (sum[i]+a[i]=k) **or** (sum[i]+a[i]+b[i+1]=k) **or** (i=n-1);
21.          **if** lower1 $\neq$ upper1 **then** sum[i] := sum[i]+d[up[i]];
22.          next_landing := (sum[i]+next=k) **or** (sum[i]+next+b[i+1]=k) **or** (i=n-1);
23.          up1[i] := up1[i-1]; up1[i-1] := i-1;   {value of up1 propagates downwards}
24.          **if** lower1=upper1 **then** down[up1[i]] := i          {case 1}
25.          **else if** next_landing **then** down[up[i]] := i          {case 2}
26.                        **else** down[up[i]] := down[i];   {case 3}
27.          **if** next_landing **then** up1[i] := i;
28.          d[i] := -d[i];
29.      **end**;
30.      **if** up_point **then begin**
31.          ii := i; i := up[i]; up[ii] := ii; up_point := **false**;   {going up}
32.      **end else begin**

```
33.         if not mark[down[i]] then solve[down[i]] := solve[i];
34.         mark[i] :=   false; i := down[i]
35.     end
36. until i=0;
37. output(a).
```

Line 8-11. Go through the cross level from node $v_i$ to next $v_i$ in the notation of Figure 1.
Line 13-35. When we hit a last child, we update data structures and decide whether to go up or go down. Most of the work is for preparing information for the opposite path.
Lines 15-16. Computation of the upper bound and lower bound of the opposite side
Line 18. solve[up[i]] has been set. Also we can use the same solve[i] for level i later.
Line 19. Declare solution points of levels at up[i] and i are known.
Line 20. If up-point is true, it means the current node has a single child, hence it is an up-point
Line 21. Prepare sum[i] for the opposite path.
Line 22. To check if the next node is a landing point.
Line 27. If next_landing is true, we set up1[i] to i.
Line 33. Prepare the solution point for level down[i] before going down.

## 7. In-Place Generation

In this section we show how to generate multiset combinations in-place in array in O(1) time per object with O(n) space. The idea is to enhance Algorithm 2 with additional data structures.

Let array "container" be the container of multiset. In Example 3, we have the initial state and next state of a and container as

```
a=(0, 0, 2, 1, 1), container=(3, 3, 4, 5)
a=(0, 1, 2, 1, 0), container=(3, 3, 4, 2)
```

That is, component 5 is out and 2 is in. If we maintain the positions of components in a stack, we can keep track of those positions. We can say a ball come from source 5 to destination 2. We express those two values by "dest" and "source". Also we prepare n stacks, stack[1], …, stack[n] for the above mentioned positions. We insert the following piece of code before line 9.

```
if d[i]>0 then begin dest:=i; source:=solve end
else begin dest:=solve[i]; source:=i end;
j:=pop(stack[source]); /* pop up from stack[source] */
push(stack[dest], j); /* push j into stack[dest] */
container[j]:=dest;
```

Initialization for the stacks, initially empty, is as follows:

```
for j:=1 to k do push(stack[container[j]], j);
```

In our example we have the following changes from left to right.

| | |
|---|---|
| container=(3, 3, 4, 5) | container=(3, 3, 4, 2) |
| dest=2, source=5 | |
| Stack[1]=empty | stack[1]=empty |
| Stack[2]=empty | stack[2]=(4) |
| Stack[3]=(1, 2) | stack[3]=(1, 2) |
| Stack[4]=(3) | stack[4]=(3) |
| Stack[5]=(4) | stack[5]=empty |

## 8. Concluding Remarks

We developed an O(1) algorithm for generating adjacent multiset combinations. Algorithm 2 is general enough to further adapt to other combinatorial objects. Recursive algorithms, which were given in Section 2, were first developed, and then converted to iterative ones based on Algorithm 1. Recursive algorithms are easier to develop since we can control the path to the leaf at the cost of O(n) time. The most difficult part is how to avoid this O(n) time by the aid of additional data structures. We partially solved the in-place generation in O(1) time spending O(n) space. A complete solution with O(1) time and O(k) space is an open problem. Also the same components are placed in an array separately in our algorithm. It is open whether we can keep them consecutively.

More formal explanation of Algorithm 2 will be a future research topic. A full Pascal programs for Algorithm 2 is attached for verification.

**Appendix**    Pascal program for generating multiset combinations

```
program ex(input,output);
label 10;
var i,i0,ii,n,k,kk,s1,count,lower,upper,lower1,upper1,next:integer;
    b,d,down,up,up1,m,sum,solve,mark:array[0..100] of integer;
    a:array[0..100] of integer;
    up_point,next_landing:boolean;
procedure out;
var i:integer;
begin count:=count+1; for i:=1 to n do write(a[i]:2); writeln
end;
function min(x,y:integer):integer;
begin if x<=y then min:=x else min:=y end;
function max(x,y:integer):integer;
begin if x>=y then max:=x else max:=y end;
begin {main}
   writeln('input k and n); readln(k,n); kk:=k;
   writeln('input m[1], ..., m[n]');
   for i:=1 to n do read(m[i]); readln;
   for i:=n downto 1 do
     if m[i]<=kk then begin a[i]:=m[i]; kk:=kk-a[i] end
     else begin a[i]:=kk; goto 10 end;
   10:
   i0:=i; b[n+1]:=0;
   for i:=n downto 1 do b[i]:=b[i+1]+m[i];
   for i:=0 to n do begin up[i]:=i; up1[i]:=i; solve[i]:=n; mark[i]:=0 end;
   sum[0]:=0; a[0]:=0;
   for i:=1 to n do sum[i]:=sum[i-1]+a[i-1];
   for i:=i0+1 to n do sum[i]:=sum[i]+1;
   for i:=1 to i0 do d[i]:=1; for i:=i0+1 to n do d[i]:=-1;
   for i:=1 to n-1 do down[i]:=n-1;
   count:=0; i:=i0;
   repeat
     out;
     lower:=max(k-b[i+1]-sum[i],0); upper:=min(k-sum[i],m[i]);
     if not((d[i]>0) and (a[i]=upper) or
             (d[i]<0) and (a[i]=lower)) then begin
       a[i]:=a[i]+d[i]; a[solve[i]]:=a[solve[i]]-d[i];
     end;
     up[i]:=i;
     if (d[i]>0) and (a[i]=upper) or (d[i]<0) and (a[i]=lower) then begin
       up[i]:=up[i-1]; up[i-1]:=i-1;
       lower1:=max(k-b[i+1]-sum[i]-d[up[i]],0);
       upper1:=min(k-sum[i]-d[up[i]],m[i]);
       if d[i]>0 then next:=upper1 else next:=lower1;
       if next<>a[i] then solve[up[i]]:=i else solve[up[i]]:=solve[i];
       mark[up[i]]:=1; mark[i]:=1;
       up_point:=(sum[i]+a[i]=k) or (sum[i]+a[i]+b[i+1]=k) or (i=n-1);
       if lower1<>upper1 then sum[i]:=sum[i]+d[up[i]];
       next_landing:=(sum[i]+next=k) or (sum[i]+next+b[i+1]=k) or (i=n-1);
```

```
            up1[i]:=up1[i-1]; up1[i-1]:=i-1;
            if lower1=upper1 then down[up1[i]]:=i
               else if next_landing then down[up[i]]:=i
                                    else down[up[i]]:=down[i];
            if next_landing then up1[i]:=i;
            d[i]:=-d[i];
      end;
   if up_point then begin
      ii:=i; i:=up[i]; up[ii]:=ii; up_point:=false;
   end else begin
      if mark[down[i]]=0 then solve[down[i]]:=solve[i];
      mark[i]:=0; i:=down[i]
   end
until i=0;
   out; writeln('count= ', count:5);
end.
```